# Building a Dataspace for Manufacturing as a Service in Factory-X

**Marco Simon**

*Technologie-Initiative SmartFactory KL e.V., Kaiserslautern, Germany – marco.simon@smartfactory.de*

**Felix Schöppenthau**

*Fraunhofer IOSB, Karlsruhe, Germany*

**Richard Kuntschke, Catharina Czech, Birgit Obst, Bertram Fuchs**

*SIEMENS AG, Munich, Germany*

**Thomas Lepper**

*DMG MORI Bielefeld GmbH, Bielefeld, Germany*

**Timo Schurek**

*Instawerk GmbH, Stuttgart, Germany*

**Steffen Currle**

*TRUMPF SE + Co. KG, Ditzingen, Germany*

**Kai Wernet**

*Matchory GmbH, Blaustein, Germany*

**Jana Pralle**

*Institut für Fertigungstechnik und Werkzeugmaschinen (IFW), Hanover, Germany*

**Pascal Rübel**

*Technologie-Initiative SmartFactory KL e.V., Kaiserslautern, Germany*

**Abstract**

*One way to solve the challenge of small and medium-sized enterprise (SME) manufacturers of acquiring sufficient orders is by joining digital Manufacturing-as-a-Service (MaaS) platforms for on-demand manufacturing. However, joining such platforms brings about new challenges such as efficient quoting handling in the face of potentially low success rates and the need for high production quality for low lot sizes. Automating the complete interaction between manufacturers and MaaS platforms, from registering the manufacturer and its capabilities to handling incoming requests and managing offers, orders, and production quality reporting, helps to overcome these challenges. Thus, the increased number of requests can be handled efficiently, and the production quality can be maintained at a high level even*

*for low lot sizes. This paper presents an architecture for automating the interaction and functional building blocks between manufacturers and MaaS platforms, along with a prototype implementation and evaluation of its effectiveness in addressing the challenges SME manufacturers are faced with.*

# 1 Introduction

The manufacturing sector faces significant challenges. In an increasingly uncertain and volatile environment, companies are under growing pressure to respond quickly and flexibly to unforeseen events. Current developments such as changing tariff policies or global warming and the associated need to reduce greenhouse gas emissions illustrate this situation. Against this background, Manufacturing-as-a-Service (MaaS) is becoming increasingly important [1]. Consequently, this use case is also reflected within the framework of Manufacturing-X. Manufacturing-X is a publicly funded initiative that aims to strengthen the resilience, sustainability, and competitiveness of the manufacturing industry through digitalization. Factory-X is the lighthouse project embedded within Manufacturing-X, aiming to establish an open, decentralized, and collaborative data ecosystem for the manufacturing industry [2]. This paper presents an implementation of a MaaS data space within the Factory-X project. MaaS data space in Factory-X addresses three central goals [2]. First, it aims to improve the visibility and transparency of manufacturers on digital marketplaces by making their capabilities discoverable and comparable. Second, it seeks to optimize the bidding process and enable competitiveness through well-matched order allocation. Third, it targets efficient and reliable order execution by automating processes and supporting high manufacturing quality even for small batch sizes. This approach specifies three central scenarios for the implementation of these goals: Supplier Capability Notification, Search, Request, Offer, and Order, and Order Execution and Quality Control. Unlike other existing approaches, this one explicitly and systematically considers the relevant stakeholders—buyers, suppliers, and particularly the on-demand manufacturing platform. In practice, data exchange and quoting are still largely manual and heterogeneous, for example through unstructured requests and non-standardized formats, which limits scalability for high request volumes on MaaS platforms. The technical implementation in this approach uses established Industry 4.0 concepts, particularly the Asset Administration Shell (AAS) and the MX-Port.

This creates consistent, manufacturer-independent collaboration that covers the entire process from order placement and execution to quality assurance, thereby going beyond the typical scope of existing solutions. The paper is structured as follows. In Section 3, we present the scenarios and the involved stakeholders using sequence diagrams. In Section 4, we present our overall demonstrator, showing how the scenarios are integrated and implemented in dedicated sub-demonstrators. In Section 5, we discuss the key findings and outline the limitations of the current implementation.

## 2   State of the Art

### 2.1   Manufacturing-X and Factory-X

Manufacturing-X is a joint initiative by industry, politics, and science to create a sovereign data space along the entire manufacturing and supply chain. The aim is to increase the resilience, sustainability, and competitiveness of industry through multilateral data use. The technological basis is formed by open standards such as AAS, OPC UA, Gaia-X, and Catena-X [3]. As part of the Manufacturing-X initiative, Factory-X is laying the technical foundation for secure, cross-company data exchange in the manufacturing and supply chain. Of the project's eleven application-oriented use cases, this paper takes a detailed look at the concept of Manufacturing as a Service within industrial data spaces [2].

### 2.2   Other Approaches for Shared Manufacturing

One approach describes how a digital MaaS ecosystem was designed for Catena-X and implemented in a demonstrator. Its focus is on matchmaking between requested and offered manufacturing capabilities. In addition, the Smart Factory Web 2.0 reference architecture is further developed by extending it with data space technologies and deliberately designing it in a generic way to support overarching governance mechanisms, for example, those defined by Gaia-X. A subset of the results developed in this project have also been used inside the MaaS use case within the current Factory-X project. Some key aspects, however, have not been addressed in Catena-X. Capacities are not explicitly considered, and the end-to-end workflow from quotation and offer preparation through ordering and execution is not covered. Moreover, processes such as cost calculation and quotation preparation have not been standardized or automated [4]. Another approach is presented in [5]. It uses the Capability, Skill, and Service (CSS) model [6] as a central concept and implements it with the AAS and the Industry 4.0 (I4.0) language [7]. The work investigates how the CSS model can be applied to enable flexible and semantically consistent collaboration within a shared manufacturing network. In this approach, two stakeholders are defined, the service provider and the service requester. Building on the bidding workflow described in [8], messages are exchanged between the actors using the I4.0 language. Since communication is carried out via Message Queuing Telemetry Transport (MQTT), an additional Registry Infrastructure Component (RIC) is introduced to manage MQTT topics and to organize the dynamic routing of messages within the bidding process. However, the approach clearly focuses on the bidding procedure. Aspects beyond this process step are not covered. An alternative approach simulates a MaaS system. Within a demonstrator, stamping and laser-cutting capabilities of resources are considered to produce, as example products, an oven door and a refrigerator door. The technical basis is provided by the AAS and the International Data Spaces (IDS) connectors. The resources are registered in an AAS registry/manager. A higher-level orchestrator receives manufacturing orders and identifies suitable assets via the registry to execute the individual production steps. Through the IDS connectors, this orchestrator is made indirectly accessible to external partners. It enables both the sharing of an AAS catalogue and the triggering of manufacturing orders across organizational boundaries. However, an explicit platform role is missing, the origin of the manufacturing order is not entirely clear, and aspects such as quality assurance are not addressed [9].

## 2.3 Asset Administration Shell (AAS)

The AAS is defined as a standardized digital representation of an asset [10] and constitutes the reference model for representing industrial assets within the Industry 4.0 framework. It provides a structured, machine-readable digital description that integrates an asset's properties, functions, and lifecycle information into a unified information model. This model is formalized in the international specification IEC 63278-1 [11]. Through its modular submodel architecture and technology-independent design principles, the AAS enables interoperable data exchange across heterogeneous systems and organizational boundaries. As a result, it forms a foundational component of industrial digital twins, supporting advanced use cases such as plug and produce integration, predictive maintenance, and cross-enterprise data sharing. By standardizing the interface between assets and digital ecosystems, AAS contributes to scalable automation and consistent data governance in modern manufacturing environments. The AAS can be categorized into three types depending on the pattern used to exchange information with it, and it is necessary to choose the right type for the respective use case [12]. Type 1 represents the AAS as a file that stores information about the asset throughout its lifecycle, for example in AASX, JSON, or XML format. This type is commonly referred to as a passive AAS. Type 2 enables online access to asset information via an API that responds to external requests. Therefore, it is called a reactive AAS. Type 3 goes one step further. A proactive AAS has its own decision-making capabilities and can initiate actions autonomously. Type 3 AAS are able to interact and collaborate with other AAS. For reaching a high level of interoperability, the AAS is based on standards and specifications. The Industrial Digital Twin Association (IDTA) plays a crucial role in providing five specifications that define different parts of the AAS. The AAS specification is structured into several parts. Part 1 (Metamodel, IDTA No. 01001-3-0-1) [13] defines the overall structure of the AAS and describes how information is exchanged between partners. Part 2 (Application Programming Interfaces, IDTA No. 01002-3-0-3) [14] specifies the API used to access AAS information and is based on HTTP/REST calls. Part 3a (Data Specification – IEC 61360, IDTA No. 01003-a-3-0-2) [15] introduces a data specification template aligned with IEC 61360 to define additional attributes that are not covered by the metamodel. Part 4 (Security, IDTA No. 01004-3-0) [16] introduces access control and security mechanisms for the AAS. Finally, Part 5 (Package File Format – AASX, IDTA No. 01005-3-0-1) [17] defines the AASX package format used for exchanging AAS data.

## 3  Manufacturing as a Service Stakeholders and Scenarios

Chapter 3 introduces the MaaS concept by first outlining the key stakeholders of the ecosystem and subsequently presenting the core interaction scenarios that structure the end-to-end processes within the Factory-X data space.

### 3.1  Stakeholders in MaaS

Figure 1 shows the main roles in MaaS and the overarching process for standardized data exchange.

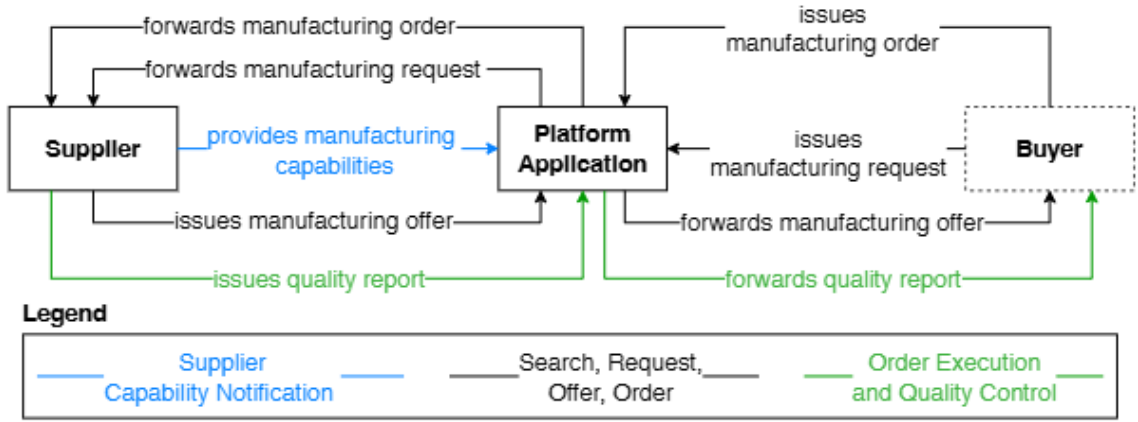

Figure 1: Roles in MaaS

- **Buyer:** A buyer is an entity, e.g., a company, buying manufacturing services from Suppliers.
- **Supplier:** A supplier is an organization, e.g., an SME, that provides manufacturing services to buyers. Suppliers can offer their services via platform applications or platforms represent themselves as supplier.
- **Platform application:** A platform application is a MaaS cloud platform that acts as a middleware between buyers and suppliers by hosting the manufacturing as a service offering of multiple suppliers and providing them to interested buyers. Suppliers can register at platform applications and publish their manufacturing capabilities there. Buyers can search for suppliers that match their manufacturing request at platform applications.

## 3.2 Scenarios in MaaS

### 3.2.1 Supplier Capability Notification (Scenario ①)

A supplier must provide comprehensive manufacturing-related information to ensure visibility on a MaaS platform. In particular, the supplier's offered manufacturing capabilities need to be systematically collected and represented in a structured, machine-interpretable format to support the following onboarding process. Figure 2 illustrates the sequence of the Supplier Capability Notification scenario. The interaction can be realized via HTTP pull or HTTP push, and the pull mechanism was validated within the Factory-X MaaS context. In this setup, authorized capability information as well as additional machine and factory data are exposed by the supplier via the MX-Port concepts through a so-called Connectivity/User Interface component. The exchanged information is represented using AAS submodel templates. On the consumer side, the platform (i.e., the Platform Application) retrieves the supplier-provided data. In the following, this scenario is referred to as Scenario ①.

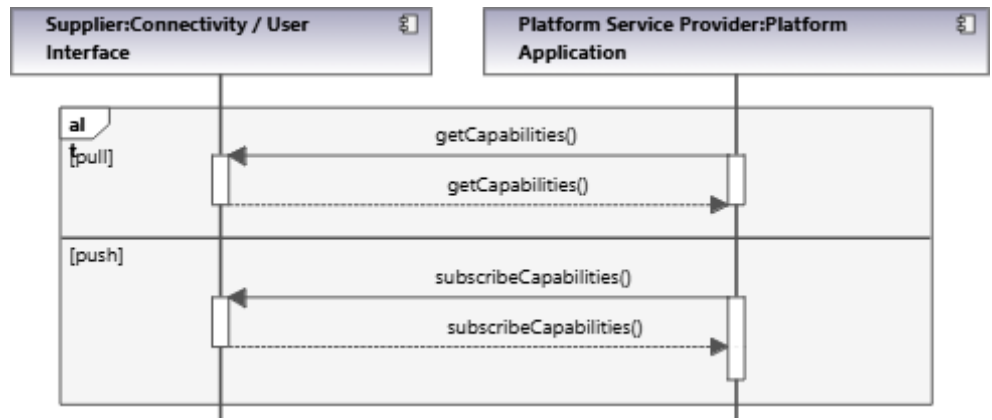

Figure 2: Supplier Capability Notification Scenario

### 3.2.2 Search, Request, Offer, and Order (Scenario ②)

The Search, Request, Offer, and Order scenario is initiated by a customer-side search for a suitable manufacturing service provider via the platform application. An overview of the complete workflow is provided in Figure 3. Based on the manufacturing capabilities required to produce the requested part or product, which are optionally derived automatically by a feature recognition service (getFeatures(in part)), the platform performs capability-based matchmaking against the capabilities published by registered suppliers. For this purpose, the platform invokes getSuppliers(in capabilities) on the matchmaking service to identify eligible suppliers.

Once suitable suppliers have been identified and the part requirements are available, the platform issues a request for quotation (RFQ). The platform submits the RFQ in a structured AAS format to the supplier via the supplier's connectivity or UI application using sendRequest (in request). On the supplier's side, the request is forwarded to the automated order processing and handled through processRequest (in request). As part of this processing, an automated costing step is triggered: the order processing calls getCostAnalysis (in features) on the automated cost analysis service to estimate the manufacturing costs based on the previously extracted features. The resulting cost assessment is then returned to the order processing, which compiles the quotation and routes the final offer back to the platform application through the connectivity or UI in AAS format. This process enables a fast and standardized request-to-offer workflow that integrates relevant manufacturing details and requirements. In the following, this scenario is referred to as Scenario ②.

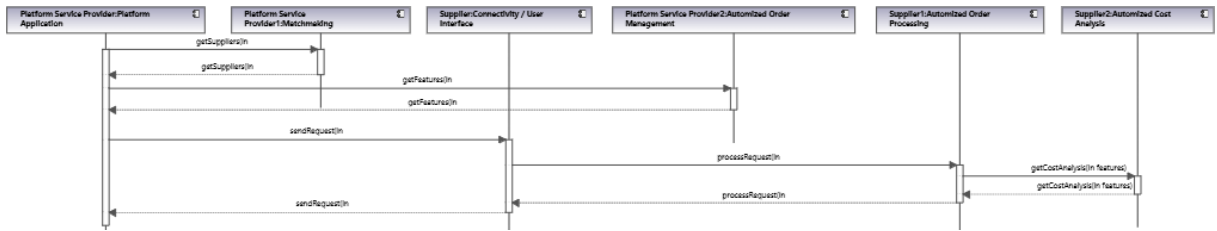

Figure 3: Search, Request, Offer, and Order scenario

### 3.2.3 Order Execution and Quality Control (Scenario ③)

The order execution and quality control scenario begins when an order is received by the manufacturer through the connectivity or user interface application. The manufacturer then initiates the production planning process, which is part of the automated order processing application shown in the sequence diagram. Key steps in this phase include feature recognition, CAM automation, and the extraction of quality requirements in order to process the order and prepare production. During machining, machine signal data is captured and transferred to a quality control application via the providedQualityData() function. The sendMeasurementReport() function then generates an AAS submodel called Quality Control for Machining, which documents both the specified requirements and the achieved results. This submodel can subsequently be sent back to the platform application and made available to the customer or buyer. In the following, this scenario is referred to as Scenario ③.

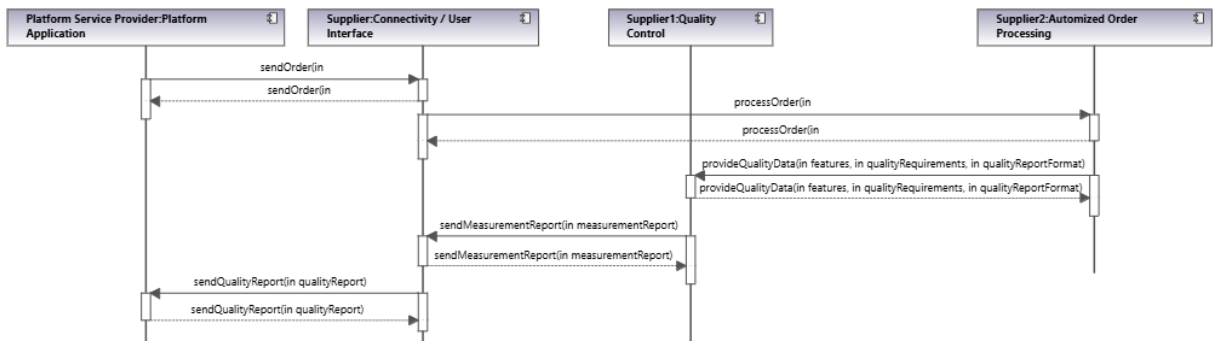

Figure 4: Order Execution and Quality Control Scenario

# 4 Implementation of Scenarios

## 4.1 Used Semantic Models

All interactions in the scenario take place through the standardized AAS Submodels listed in Table 1.

Table 1: Overview of AAS Submodel Standards, Versions, Affiliations, and Scenario Coverage

| Standard | Ver. | Affiliation | Scenario |
| --- | --- | --- | --- |
| Digital Nameplate for Industrial Equipment | 3.0 | Company, Factory, Machine, Product | ①, ②, ③ |
| Capability Description | 1.0 | Factory, Machine, Product | ① |
| Purchase Order | 1.0 | Product | ② |
| Quality Control for Machining | 1.0 | Product | ③ |
| Handover Documentation | 2.0 | Factory, Machine, Product | ②, ③ |
| Generic Frame for Technical Data for Industrial Equipment in Manufacturing | 2.0 | Company, Factory, Machine, Product | ②, ③ |
| Contact Information | 1.0 | Company, Factory, Machine | optional |
| Data Model for Asset Location | 1.0 | Factory, Machine | optional |
| Asset Interface Description | 1.0 | Asset | optional |
| Asset Interfaces Mapping Configuration | 1.0 | Asset | optional |
| Time Series Data | 1.1 | Machine | optional |
| Hierarchical Structures enabling Bills of Material | 1.1 | Company, Factory, Machine, Product | optional |
| Production Calendar | 1.0 | Machine | optional |
| Carbon Footprint | 1.0 | Product | optional |

## 4.2 Data Exchange via MX-Ports

The MX-Port [18] concept, as a core element for application connectivity, comprises a modular stack of five layers, each serving a specific purpose. The first layer, the adapter, adapts internal interfaces to the requirements of the MX-Port. The second layer, the converter, transforms internal data formats into a standardized MX-Port data format, such as AAS. The third layer, the gate, provides a standardized interface for data access. The fourth layer, access and usage control, ensures secure data access and usage. Finally, the fifth layer, discovery, enables the discovery of entities and data resources within the data space. Section 4.3.2 shows an example implementation of MX-Ports.

## 4.3 Comprehensive Manufacturing as a Service Demonstrator

To highlight the advantages of a common data space for handling complex information flows among the different actors within the Manufacturing-as-a-Service ecosystem, a comprehensive demonstrator has been developed. Figure 5 illustrates an overview of the selected focus areas, aligned with the three identified scenarios presented in Section 3. Due to the diversity of the underlying IT infrastructure, Figure 5 does not provide a complete overview but instead showcases the main parts of the demonstrator. It addresses the factory operator or supplier and the platform application as the main stakeholders.

The basis for easy and convenient access to digital marketplaces is a digital twin of the production environment, in which manufacturing resources and their capabilities are described using AAS submodels. To enable the automated description of manufacturing resources, a capability modeler, introduced in Section 4.3.1, is used. Based on these standardized AAS models for manufacturing capability description, the connectivity to the Factory-X data space is presented in Section 4.3.2. In this context, information is exchanged between the factory connector and the data space consumer on the platform side in accordance with the principles of the Factory-X data space, using the MX-Port implementations Leo and Hercules. To complete the objective of streamlined access to manufacturing marketplaces, the supplier onboarding process, including capability notification to the platforms, is described in Section 4.3.3. The Search, Request, Offer, and Order scenario begins with the configuration of a buyer's request, which is explained in more detail in Section 4.3.5. On the platform side, the identification of suitable suppliers is supported by a capability-based matchmaking process. This process can be further enhanced by geometrical feature recognition, described in Section 4.3.6, which is used to derive the required capabilities and match them with the offered capabilities of registered manufacturers. On the supplier side, an automated bidding process helps reduce manual effort. For suppliers, it is particularly important to provide fast and realistic cost estimations. To support this preliminary cost estimation, geometrical feature recognition can also be applied at this stage. Subsequently, Computer-Aided Manufacturing (CAM) automation, presented in Section 4.3.8, supports production, while quality assurance is carried out in an automated manner, as discussed in Section 4.3.9. Across all parts of the demonstrator, AAS submodels are used to improve connectivity and semantic interoperability between the different stakeholders.

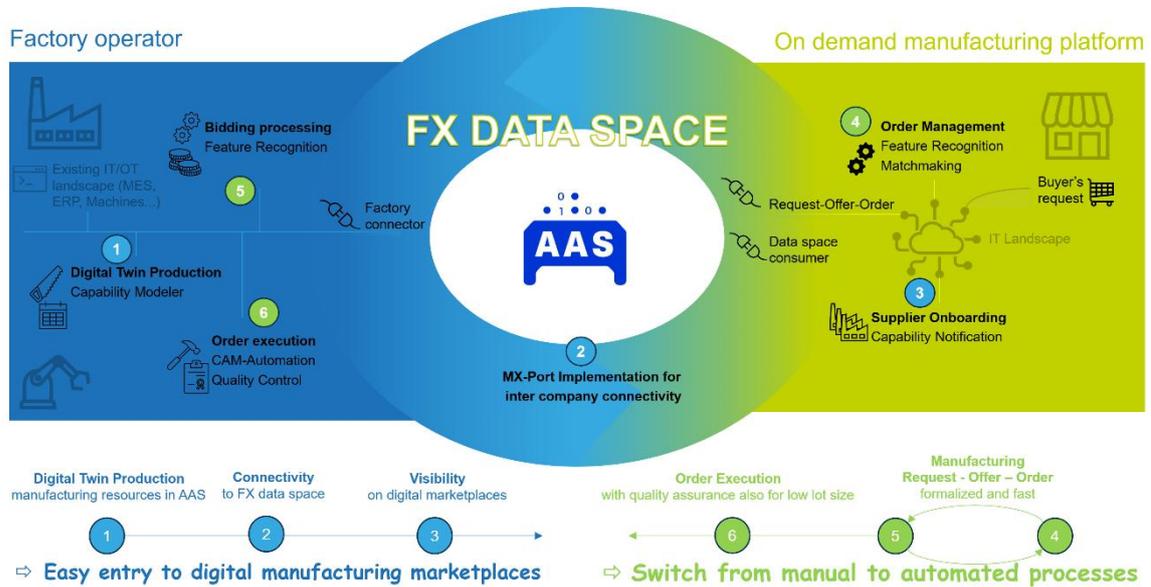

*Figure 5: Comprehensive manufacturing as a service demonstrator*

### 4.3.1 Capability Modeler

Capability Modeler: To transform the description of production capabilities from a manual into an automated task, a capability modeler has been developed. Capabilities are derived from historical production and feature data retrieved from a self-defined manufacturing feature submodel within an AAS repository, where all relevant feature information is stored. This feature information serves as input to the capability modeler and is processed to determine capabilities in a standardized representation. In addition, free-text sources such as written documentation of machine capabilities can also be considered and transformed accordingly. Once the submodel Capability Description is generated, it is written back to the AAS repository and assigned to the corresponding product, machine, or factory AAS, enabling the structured and machine-interpretable storage of manufacturing-relevant capabilities. The implementation is based on the BaSyx Python SDK, which provides the AAS metamodel. Building on this foundation, dedicated classes and functions have been developed to automatically generate the submodel. The functionality is provided via an application programming interface (API) and is accessible through a graphical user interface for structured input handling. Additionally, an MCP server enables the transformation of free-text input into formalized capability descriptions.

### 4.3.2 MX-Port Implementation for Inter-Company Connectivity

Next, the connectivity between the manufacturer and the platform, following the principles of the Factory-X data space, is described. Among other things, this connectivity concept can be used to publish the previously identified manufacturing capabilities to platforms. As highlighted in Figure 6, multiple MX-Port variants can be used for standardized data exchange between the main MaaS roles, namely the supplier and the platform application. Both stakeholders can act as data consumers as well as data providers. Within the Factory-X MaaS context, data exchange between a data consumer and a data provider was implemented and successfully tested using MX-Port Hercules and MX-Port Leo.

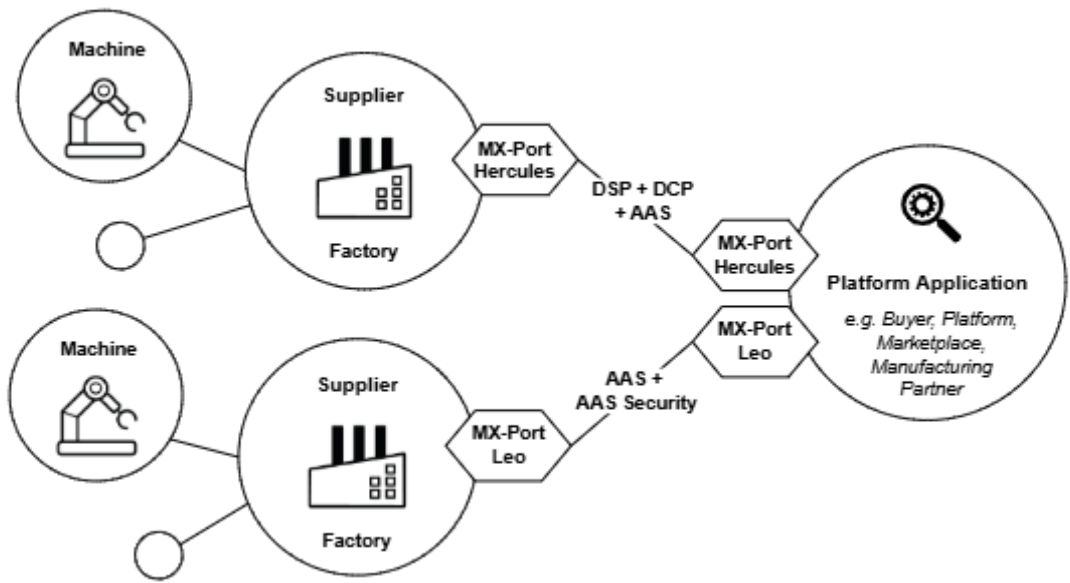

Figure 6: Overview Stakeholder

*Architecture example for data exchange via MX-Port Hercules.*

Figure 7 shows a concrete implementation of the MX-Port Hercules given an exemplary set of software components.

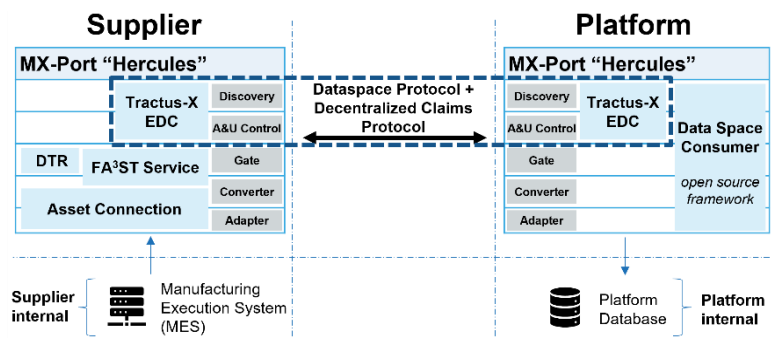

Figure 7: MX-Port Hercules Implementation Example

From the supplier's perspective, the objective is to automatically integrate Manufacturing Execution System (MES) data into AAS submodels using an asset connection, realized through the Asset Connection feature of the FA³ST Service. The AAS repository, implemented with the FA³ST Service, provides the AASs via a standardized REST API. In addition, specific access control rules for individual AAS resources can be defined within the AAS repository as part of AAS Security. The FA³ST Service also automatically registers the AASs in the Digital Twin Registry (DTR), which is realized by the FA³ST Registry. The DTR serves as a directory service that provides information about which data is available in which AAS repositories, as well as the data space endpoints that must be addressed by a consumer to access the respective AAS data. The Tractus-X EDC exposes both the DTR and the AAS repository as datasets within the data space, so that these services do not need to be made directly accessible. Data is exchanged via the EDC after successful negotiation of the provided contract and a subsequent authorized data request by the consumer. The consumer likewise requires a Tractus-X EDC in order to negotiate a contract with the provider EDC. To improve ease of use and facilitate adoption on the consumer side in this use case, an open-source software component called Data Space Consumer was released as part of Factory-X MaaS. This component automates the sequence of requests to a provider that is necessary to retrieve AAS data. Moreover, the automated request mechanism of the Data Space Consumer is independent of the specific use case and complies with the Factory-X Architecture Decision Records. Out of the box, the Data Space Consumer provides functionality for the discovery, access and usage control, gate, converter, and adapter layers, including components such as the faaast-gate-extension.

*Architecture example for data exchange via MX-Port Leo*

Figure 8 depicts a concrete example implementation of the MX-Port configuration Leo using an exemplary set of software components such as the AASPE Server and the Data Space Consumer.

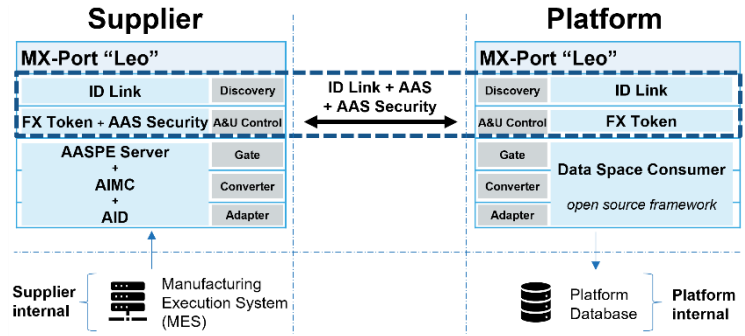

*Figure 8: MX-Port Leo Implementation Example*

In this implementation, the AASPE Server is used on the supplier side together with the Asset Interface Description (AID) and Asset Interfaces Mapping Configuration (AIMC) submodel templates to implement the adapter, converter, and gate layers of the MX-Port. The access and usage control layer is implemented using the FX Token for authentication, as well as access rules defined through AAS Security for authorization, as described in the chapter "Data Exchange Steps via MX-Port Leo" above. The discovery layer is implemented using ID Link. On the platform side, the Data Space Consumer is used again, this time with extensions that implement the access and usage control layer and the discovery layer of MX-Port Leo using the FX Token and ID Link, respectively.

### 4.3.3 Capability Notification

In the previous section, intercompany connectivity was described in detail, with the roles of data consumer and data provider being assumed by suppliers, platforms, or other service providers. This concept also enables suppliers to automatically share manufacturing and supply chain data. A demonstrator for capability notification therefore uses the MX-Port concept to showcase the supplier capability notification scenario introduced in Section 3.2.1. Data consumers, such as digital platforms and marketplaces, use the provided capability data to add suitable manufacturers to their databases for capability-based matchmaking when a buyer searches for the most suitable manufacturer. The now standardized format for capability description allows these processes to be automated much more effectively, as illustrated in Figure 2.

### 4.3.4 Matchmaking

Access to a standardized form of supplier information can be extremely helpful, particularly in the matchmaking process, as it helps minimize semantic interoperability problems between the database and the query system. In particular, the structured recording of capability data can lead to significantly improved search results.

### *4.3.5 Request*

Based on the scenario described in Section 3.2.2, the Search-Request-Offer-Order workflow demonstrator illustrates the process from the customer's perspective, starting with obtaining an offer for parts to be manufactured, then translating this offer into an order, or directly creating an open order based on the estimated costs within the MaaS environment. In this step, the project AAS is created based on the standardized Purchase Order AAS submodel and enriched with all available product information, which is then complete and formalized to support subsequent automated processing. In addition, the results of feature recognition are added as part of the Technical Data AAS submodel, enabling potentially automated processes on the manufacturer side to better assess their ability to carry out the project. Once the order is accepted by both parties, the process proceeds to order execution.

### *4.3.6 Feature Recognition*

Feature recognition can be used at different stages of the process. First, it can support the search for a suitable manufacturer by identifying the capabilities required to produce a given part. Second, it can assist the manufacturer during bidding and production planning. To enable preliminary cost estimation and additional automation services in the MaaS environment, machine learning (ML)-based manufacturing feature recognition is applied to the CAD model provided by the customer. The ML model identifies both the semantics of manufacturing-relevant feature classes, such as holes, pockets, and slots, and the corresponding feature instances, that is, the specific occurrences of these features within the part. The resulting feature set is written into the standardized Technical Data AAS submodel and linked to the underlying CAD topology using CAD part face references. For each feature instance, characteristic parameters such as diameter, depth, radius, or the number of feature faces are stored in a machine-interpretable form, enabling direct use by downstream processes such as feature-based cost estimation, capability matchmaking, CAM automation, and inspection planning. In addition to instance-specific data, aggregated AAS elements can also be provided, including feature counts per class, estimated removal volume, minimum required tool diameter, and feature-specific confidence values to support automated decision-making.

### 4.3.7 Cost Estimation

To further reduce manual effort during quotation and order preparation, the demonstrator supports an automated cost estimation process based on the order data generated within the Purchase Order submodel. The order is transferred to the contract manufacturer, who determines a manufacturing price for the component. This estimation is based on the feature instances identified in the Technical Data AAS submodel. These features are enriched with additional attributes such as geometric dimensions and their positions on the component. Based on this enhanced feature description and the available machine capabilities, suitable machines and tools are assigned. To estimate process times, a similarity-based algorithm is applied that compares features at a geometric level. When similar features are identified, historical data is used to scale the corresponding process times. Using the predicted process times and the machine hourly rates stored in the Technical Data AAS submodel of the machines, feature-level costs are calculated. These costs are then stored within the respective feature AAS, making them available for downstream processes. By aggregating the costs across all features, the contract manufacturer can ultimately derive a total price for the component.

### 4.3.8 CAM-Automation

To further reduce manual effort during order preparation, the demonstrator supports an automated CAM planning step based on the extracted feature instances and their semantic classes. The model predicts suitable CAM operation types and tool candidates for each feature, resulting in a machine-interpretable preliminary process plan. Based on the assigned operations and tools, quantitative indicators such as material removal volumes, tool-change estimates, and setup-related complexity measures can be derived. These results support downstream tasks including capability checks depending on machine or tool constraints, lead-time estimation, and inspection planning. The outputs can be integrated into the product AAS by adding CAM-related elements either to the Technical Data submodel or to the Handover Documentation.

### 4.3.9 Quality Control

Besides CAM planning, automated machining quality estimation based on production data represents another key focus area within overall order execution at the manufacturer. The demonstrator is based on the standardized Quality Control for Machining AAS submodel, which has been further refined with additional input from the MaaS Factory-X project. This AAS submodel is used to formally describe specific quality requirements before production and to document the corresponding results after production. This includes integrating the Product Manufacturing Information (PMI) directly into the AAS and subsequently linking it to the machine controller. A crucial aspect at this stage is the setup of the Quality Fingerprint, which serves as a baseline for process monitoring. The part is then machined while, at the same time, the quality fingerprint is evaluated in parallel with the process, enabling real-time monitoring. Statistical metrics such as skewness and kurtosis of the fingerprint are used to assess whether any abnormalities occurred during the machining process. These values are then compared against a threshold, which is either provided by an experienced machine operator or derived by comparing the initial production samples with actual measurement data. The resulting quality estimation is written into the corresponding AAS submodel linked to the product AAS, thereby forming a quality report that can be provided to both the manufacturing platform and the buyer.

# 5 Discussion

The MaaS in Factory-X approach was broken down into three scenarios and embedded in a comprehensive overall demonstrator. This demonstrator shows that the scenarios can be technically integrated and mapped end to end using the listed semantic models. Using AAS submodels provides a shared semantic layer, replaces heterogeneous data formats, and enables a machine-readable interface between stakeholders. However, mapping company-specific data to standardized submodels is not always trivial. In particular, the required level of detail is not always clear, as capabilities can be described at different levels of granularity, which can lead to inconsistencies. The MX-Port supports data exchange between buyer, supplier, and platform and allows the controlled sharing of information across companies. Based on this setup, automation can accelerate RFQ handling, make quality requirements more transparent, and support faster offer preparation. This can help SMEs cope with high RFQ volumes and low success rates, while also supporting quality assurance for small batch sizes. At the same time, the benefits depend strongly on data quality and on the consistent maintenance of the required information in the submodels. Initial effort for onboarding, data mapping, and ongoing data maintenance must also be considered.

# 6 Conclusion and Future Work

This contribution presents an end-to-end perspective on Manufacturing-as-a-Service (MaaS). The process chain is automated throughout, ranging from supplier capability notification and the search-request-offer workflow to order execution and quality assurance. All relevant stakeholders, namely the buyer, supplier, and platform, are explicitly considered. The MX-Port ensures data exchange between these actors and enables a sovereign and interoperable transfer of information. Building on this foundation, the applied AAS submodels support a consistent and machine-readable representation of the required data and process steps. Based on this approach, MaaS can in the future be extended to cover the entire product value chain. While the presented implementation focuses on machine tools, additional manufacturing steps such as assembly, painting, or other process chains could be integrated and transferred into an end-to-end supply chain. Within such a network, the anonymized exchange of process data, for example from machining processes, becomes feasible. This could provide the basis for a learning system that optimizes typical manufacturing steps in a data-driven manner and benefits from the aggregated experience of many participants. In addition, it appears promising to investigate to what extent large language models can support the automated transformation of heterogeneous data into AAS-compliant submodels. To further substantiate the practical applicability of the approach, a broader evaluation involving multiple real SMEs is recommended. Such an evaluation could examine scalability, bidding success rates, and onboarding effort, while process studies could provide additional insights into acceptance by suppliers and platforms. Finally, the manufacturing of IP-critical products requires robust access control, usage restrictions, and mechanisms to detect and sanction contractual violations in a traceable manner.

## Acknowledgment

This work was partially supported by the German Ministry for Economic Affairs and Climate Action by the project "Aufbau eines Datenraums für die ausrüstende Industrie – die Ausrüster von Fabriken weltweit" (Factory-X) under grant 13MX001ZU.

## References

[1] A.-L. Andersen, T. D. Brunoe, E. B. Worup et al., "Manufacturing-as-a-Service (MaaS) to Increase Value Chain Resilience and Circularity," IFAC-PapersOnLine, vol. 59, no. 10, pp. 464–469, 2025.

[2] "FACTORY-X: The digital ecosystem." https://factory-x.org/

[3] plattform-i40, "Initiative Manufacturing-X," 2025. https://www.plattform-i40.de/

[4] F. Schoppenthau, F. Patzer, B. Schnebel et al., "Building a Digital Manufacturing as a Service Ecosystem for Catena-X," Sensors, vol. 23, no. 17, p. 7396, Aug. 2023.

[5] M. Simon, C. Urban, M. Winter, and A. Kocher, "Realization of a Shared Manufacturing Network using Capabilities, Skills and Services," in 2025 IEEE 30th ETFA, Porto, Portugal, pp. 1–8.

[6] A. Kocher, A. Belyaev, J. Hermann et al., "A reference model for common understanding of capabilities and skills in manufacturing," at - Automatisierungstechnik, vol. 71, no. 2, pp. 94–104, Feb. 2023.

[7] VDI, VDI/VDE 2193 Blatt 1 - Sprache für I4.0-Komponenten - Struktur von Nachrichten, Apr. 2020.

[8] VDI, VDI/VDE 2193 Blatt 2 - Language for I4.0 components - Interaction protocol for bidding procedures, Jan. 2020.

[9] M. A. Inigo, J. Legaristi, F. Larrinaga et al., "Towards Standardized Manufacturing as a Service through Asset Administration Shell and International Data Spaces Connectors," in IECON 2022, Brussels, Belgium, pp. 1–6.

[10] IEC, "IEC 63278-1:2023 Asset Administration Shell for industrial applications - Part 1," International Standard, 2023.

[11] IDTA, Ed., Specification of the Asset Administration Shell – Part 1: Metamodel. IDTA, 2023.

[12] Plattform Industrie 4.0, "Verwaltungsschale in der Praxis," BMWi, Berlin, Diskussionspapier, 2020.

[13] IDTA, "Specification of the AAS - Part 1: Metamodel (IDTA-01001)," Specification, 2025.

[14] IDTA, "Specification of the AAS - Part 2: Application Programming Interfaces (IDTA-01002)," Specification, 2025.

[15] IDTA, "Specification of the AAS - Part 3a: Data Specification Template - IEC 61360 (IDTA-01003a)," Specification, 2025.

[16] IDTA, "Specification of the AAS - Part 4: Security (IDTA-01004)," Specification, 2025.

[17] IDTA, "Specification of the AAS - Part 5: Package File Format (AASX) (IDTA-01005)," Specification, 2025.

[18] "MX-Port Concept – Enable data sharing across industries." https://factory-x.org/wp-content/uploads/MX-Port-Concept-V1.10.pdf